\documentclass[twocolumn,floatfix,amsmath,amssymb]{revtex4}
\usepackage{graphicx}
\usepackage{dcolumn}
\usepackage{color}

\newcommand{\be}{\begin{equation}}
\newcommand{\ee}{\end{equation}}

 \newcommand{\bb}{\mathbf{b}}

\newcommand{\bv}{\mathbf{v}}

\begin{document}
\title{Rapid directional alignment of velocity and magnetic field in 
magnetohydrodynamic turbulence}

\author{W. H. Matthaeus$^1$, A. Pouquet$^2$,
P. D. Mininni$^{2,3}$, P. Dmitruk$^1$, and B. Breech$^1$}
\affiliation{$1$ Bartol Research Institute and Department of Physics 
    and Astronomy,
    University of Delaware, Newark DE 18716, U.S.A. \\
$2$ NCAR, P.O. Box 3000, Boulder, Colorado 80307-3000, U.S.A. \\
$3$ Departamento de F\'\i sica, Facultad de Ciencias Exactas y 
    Naturales, Universidad de Buenos Aires, Ciudad Universitaria, 1428 
    Buenos Aires, Argentina.
}
\date{\today}

\begin{abstract}
We show that local directional alignment of the velocity 
and magnetic field fluctuations occurs rapidly 
in magnetohydrodynamics for a variety of parameters.
This is observed both in direct numerical simulations and in
solar wind data. 
The phenomenon is due to an alignment between the magnetic field and either
pressure gradients or shear-associated kinetic energy gradients.
A similar alignment, of velocity and vorticity, occurs in the Navier Stokes fluid case.
This may be the most rapid and robust
relaxation process in turbulent flows, and leads to a local weakening of the nonlinear terms in the small scale vorticity  and current structures where alignment takes place.
\end{abstract}

\maketitle

In magnetohydrodynamic (MHD) turbulence, 
the fluctuating magnetic field $\bf b$ 
and velocity fluctuation $\bf v$ 
enter on nearly equal footing.
One consequence
is that the nonlinear MHD
equations are in effect linearized 
when the fluctuation components of the magnetic field (in Alfv\'en 
speed units) are everywhere 
equal (or opposite) to the velocity field. 
Such  ``Alfv\'enic'' states have long been thought to be favored in 
relaxation processes \cite{Woltjer58a}. 
Near Alfv\'enic states are observed in the 
solar wind plasma \cite{BelcherDavis71}, 
mostly in the inner heliosphere \cite{BavassanoEA82b}.
Global evolution towards the Alfv\'enic state,
or ``dynamic alignment,'' 
\cite{DobrowolnyEA80-prl},
when it occurs, requires many nonlinear eddy turnover times.
Here we describe a related, more rapid relaxation process, 
in which local, directional near-alignment of $\bf v$ and $\bf b$
emerges in less than one turnover time, 
for a wide variety of turbulence parameters. This process
need not be associated with global alignment, but rather 
occurs independently in numerous localized patches. 

Dynamic alignment competes with other MHD 
relaxation processes as shown in simulations \cite{TingEA86}
where, for some parameters, it 
does not occur, or is incompletely realized. 
Solar wind observations also show that the degree of Alfv\'enicity
tends to decrease with increasing heliocentric distance, in apparent contradiction
to the dynamic alignment principle.
There are suggestions 
that directional alignment (a necessary condition
for global dynamic alignment)
may be more ubiquitous. 
When MHD relaxation is 
described by a
constrained energy minimization principle \cite{TingEA86},
cross helicity (Alfv\'enicity) and a magnetic invariant
(helicity in three dimensions -- 3D; or mean square flux function in two dimensions -- 2D) are held constant, while
energy is minimized. The emergent Euler-Lagrange equations
predict final states, in both 2D and 3D, 
and for all parameters, in which 
$\bf v$ and $\bf  b$
are directionally aligned or anti-aligned. 
This theory is reasonably 
well confirmed by direct numerical simulations.
Evidently, in the long time limit 
for decaying MHD turbulence, pointwise directional alignment is
obtained more easily than is the global Alfv\'enic state. 
Below we show, using MHD numerical experiments,
that local directional alignment is even more robust, 
occurs more rapidly, and appears locally, in patches.
The distribution of alignment angle 
in the solar wind is shown to be consistent with this picture.
This rapid relaxation can be understood by an elementary examination of
the MHD equations. While these features appear not to have been fully recognized previously,
the situation is in fact analogous to the local emergence of Beltrami flows 
\cite{PelzEA85} in hydrodynamics.

{\it MHD and Alfv\'enic states.}
In familiar dimensionless (``Alfv\'enic'') units, the 
equations of incompressible MHD  
are
\begin{eqnarray}
\frac{\partial {\bf v}}{\partial t} + {\bf v \cdot \nabla v} &=& 
    -{\bf \nabla} {\mathcal P} + {\bf j \times b} + \nu \nabla^2 {\bf v}
    \label{eq:momentum} \\
\frac{\partial {\bf b}}{\partial t} + {\bf v \cdot \nabla b} &=&
    {\bf b \cdot \nabla v} + \eta \nabla^2 {\bf b} ,
    \label{eq:induc}
\end{eqnarray}
with ${\bf \nabla \cdot v} = {\bf \nabla \cdot b} = 0$. Here, ${\bf v}$ 
is the velocity field, and ${\bf b}$ is the magnetic field, related to 
the electric current density ${\bf j}$ by ${\bf \nabla \times b} = {\bf j}$; 
${\mathcal P}$ is the pressure. The viscosity $\nu$ and magnetic 
diffusivity $\eta$ define mechanical and magnetic 
Reynolds numbers respectively as $R_V = LU/\nu$ and $R_M = LU/\eta$. 
Here $U = \left<|{\bf v}|^2\right>^{1/2}$, 
with $\langle \dots \rangle $ denoting a spatial average, 
and $L$ is a length scale associated with 
the large-scale flow (integral length scale).
The total energy $E = E_v + E_b = \langle |{\bf v} |^2 + |{\bf b}|^2 \rangle/2$, 
the cross helicity $H_c = \langle {\bf v} \cdot {\bf b} \rangle$, and the 
magnetic helicity $H_m = \langle {\bf a} \cdot {\bf b} \rangle$ are 
ideal ($\nu = \eta = 0$) invariants in 3D. Here 
${\bf b} = \nabla \times {\bf a}$.
Dimensionless measures of the helicities
are $\sigma_c = 2H_c/E$ and $\sigma_m = (E_L-E_R)/E_b$, 
where $E_L$ and $E_R$ are magnetic energy in left- and right-handed magnetic
polarizations, respectively.

{\it Simulations.}
We consider several sets of simulations (see Table I), 
in which the MHD equations are integrated in a 
spatially periodic domain of side $2\pi$, 
using a second order Runge-Kutta method, and 
either $2/3$-rule dealiased \cite{OrszagPatterson72},
or non-dealiased pseudospectral methods. 
All runs freely decay in time, with no external forcing.
 
The type labelled RAN are $128^3$ incompressible runs, with random broadband
initial conditions. Four cases are distinguished 
by their values of 
$\sigma_m$
and $\sigma_c $, spanning a range of possibilities for relaxation
starting from a fully random state.

We also employ two other types of initial conditions in which helicities are controlled and the fields are more ordered. 
OT runs are a generalization 
of the 2D Orszag-Tang (OT) vortex \cite{OrszagTang79}, 
a standard large scale initial 
condition for MHD turbulence. 
In our OT case, initially energies $E_v=E_b=2$, 
$\sigma_c \approx 0.4$, and $\sigma_m \approx 0$.
Another set or runs labelled ABC
consists of a parameterized large scale helical
flow, an uncorrelated and helical large scale magnetic field,
and added noise with energy spectra $\sim k^{-3}\exp [-2(k/k_0)]^2$ at $t=0$, with $k_0=N/6$ 
\cite{1536}.
These runs have $E_v = E_b = 0.5$, $\sigma_c \approx 1 \times 10^{-4}$, and 
$\sigma_m \approx 0.5 $, while numerical resolution and Reynolds numbers 
vary (see Table I). 
Finally we analyze a small spatial region near a current 
sheet in a very high Reynolds number ABC simulation, ABC4 in the table.

\begin{table}
\caption{\label{table} Parameters in the MHD simulations shown in 
the figures. 
RAN,  OT, and ABC are described in the text. 
$N$ is the resolution, $\nu$ and $\eta$ are respectively the kinematic 
viscosity and magnetic diffusivity, and $\sigma_c$ and $\sigma_m$ reffer 
to the normalized cross and magnetic helicities defined in the text. For 
the 2D run, $\sigma_m$ is based on the mean-square flux function.}
\begin{ruledtabular}
\begin{tabular}{ccccc}
Run & $N^3$&$\nu=\eta$ & $\sigma_c$ & $\sigma_m$ \\
\hline
RAN1   & $128^3$ &   $2.5 \times 10^{-3}$ &  0 & 0 \\
RAN2   & $128^3$ &  $2.5 \times 10^{-3}$ &  0.5 & 0 \\
RAN3   & $128^3$ & $2. 5 \times 10^{-3}$ &  0 & 0.5 \\
RAN4   & $128^3$  & $ 2.5 \times 10^{-3}$  &  0.5 & 0.5 \\
\hline
OT1   & $128^3$  & $5\times 10^{-3}$  & 0.4 & 0   \\
OT2   & $256^3$  & $1.5\times 10^{-3}$  & 0.4 & 0 \\
OT3   & $512^3$  & $7.5\times 10^{-4}$ & 0.4 & 0  \\
\hline
ABC1  & $128^3$  & $3\times 10^{-3}$   &  0 & 0.5  \\
ABC2  & $256^3$  & $1.25\times 10^{-3}$  &  0 & 0.5 \\
ABC3  & $512^3$  & $6\times 10^{-4}$   &  0 & 0.5  \\
ABC4   & $1536^3$ & $2 \times 10^{-4}$ &  0 & 0.5 \\
\hline
2D  &  $1024^2 $ &  $2.5 \times 10^{-4}$ &  0 & 0  \\
\end{tabular} \end{ruledtabular} \end{table}

{\it Probability density functions.}
Our main diagnostics are 
probability density functions (pdfs) of the 
local cosine of the angle $\theta$ 
between $\bv$ and $\bb$
\begin{equation}
\cos{\theta} = \cos(\bv,\bb) = \frac{\bv \cdot \bb}{|\bv||\bb|} 
\end{equation}
which are computed for each run. 

\begin{figure}[tbh]
\centerline{\includegraphics[width=7.5cm]{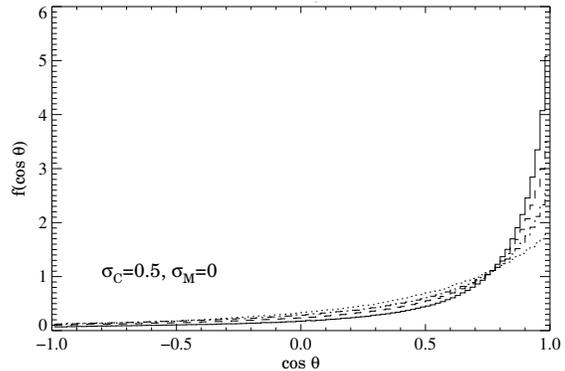}}
\caption{Pdfs of $\cos \theta$ 
for initial normalized cross helicity $\sigma_c = 0.5$ for Run RAN2. 
Global normalized cross helicity is 0.24 at t=2.
Different lines are for different times (see text).
}
\label{fig1}
\end{figure}

\begin{figure}[htb]
\centerline{\includegraphics[width=7.5cm]{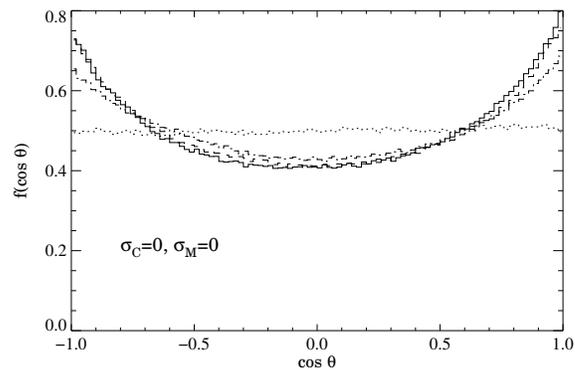}}
\caption{Pdfs of $\cos \theta$ at times
t=0 (dotted), 0.5 (dash-dotted), 1.0 (dashed), 2 (solid) 
in a 3D simulation 
$\sigma_c \approx  0$ and $\sigma_m \approx 0$ (Run RAN1).
The initial distribution is flat. 
}
\label{fig2}
\end{figure}

The distribution function for RAN2
is shown Fig(\ref{fig1}), at times $t= 0, 0.5, 1.0$ and $2.0$. 
These distributions are highly peaked near $\cos \theta \approx 1$, 
much more so than would be needed to account for the cross helicity
which is initially $\sigma_c \approx  0.5$, 
decaying to $\sigma_c = 0.24$ at $t=2.0$.
The more peaked curves are for the progressively later times. 
The results for RAN1, having no helicities, are shown in Fig. \ref{fig2}.
Now, the distributions are 
suppressed near  $\cos \theta \approx 0$ and 
strongly peaked near $\cos \theta \approx \pm 1$ 
indicating an enhanced probability of magnetic and velocity field 
being very nearly aligned or antialigned. 
Enhanced directional alignment occurs even when 
the globally averaged
cross helicity is approximately zero. 
We do not show results for RAN3 and RAN4, with $\sigma_m \approx 0.5$,
as the distributions 
are almost indistinguishable 
from the corresponding case with $\sigma_m \approx 0$.

\begin{figure}
\centerline{\includegraphics[width=7.5cm]{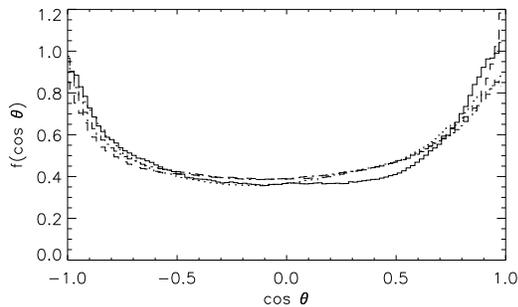}}
\caption{Pdfs of $\cos \theta$ in runs ABC1 (solid), ABC2 (dotted), 
    and ABC3 (dashed) at the peak of energy dissipation and ABC4 
    (dash-dotted).}
\label{fig4}
\end{figure}

The pdfs in the OT runs (not shown)
are asymmetric and strongly peaked at $\cos \theta \approx 1$, as in 
the RAN2 and RAN4 cases. 
For the ABC runs, with no net cross-helicity, the pdfs peak at 
$\cos \theta \approx \pm 1$ after less than half a turnover time, 
following the pattern of the RAN runs. Figure \ref{fig4} shows the 
pdfs from the ABC runs at the peak of dissipation ($t \approx 4$) for 
different Reynolds numbers.

This local alignment process is fast in all cases, with substantial 
and apparently nearly saturated alignment occurring
in less than one large 
scale turnover time. As stated above, no clear dependence with the 
Reynolds numbers is seen when we compare cases ABC1-4.

\begin{figure}
\centerline{\includegraphics[width=7.5cm]{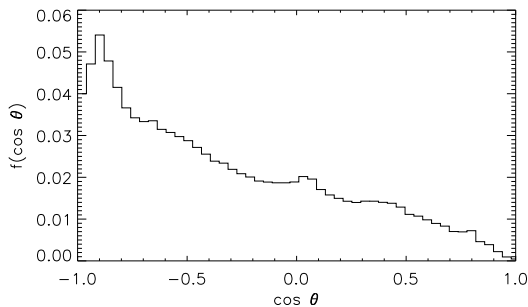}}
\caption{Pdfs of $\cos \theta$ in the vicinity of a current sheet 
    (sub-volume of $150^3$ grid points) in a $1536^3$ simulation with ABC plus noise 
    initial conditions (run ABC4).}
\label{fig5}
\end{figure}

When pdfs of $\cos \theta$ are computed in the vicinity of 
a cluster of strong current sheets, or in regions of strong shear in 
the magnetic field (run ABC4), an only slightly different result is 
obtained (Fig. \ref{fig5}). Inside the current sheet, the magnetic 
and velocity field are strongly antialigned (which gives the peak near 
$-1$), and the pdf is linear. As larger subvolumes surrounding 
the current sheet are considered, 
or at later times when current sheets accumulate and interact, 
and thus more current sheets with different 
alignments are integrated in the sub-volume, 
the pdf converges towards
the form seen in Fig. (\ref{fig4}) for the whole flow.

{\it Physics of alignment.}
Why does local alignment take place in these simulations? And why is 
it so fast?  Manipulating the MHD equations,  
Eqs. (\ref{eq:momentum}) and  (\ref{eq:induc})  
in the ideal case ($\nu = \eta = 0$),  one finds 
the equation for evolution 
of the local cross helicity:
\begin{equation}
\frac{\partial (\bv \cdot \bb)}{\partial t} + \bv \cdot \nabla (\bv \cdot \bb) 
    = \bb \cdot \nabla \frac{\bv^2}{2} - \bb \cdot \nabla {\cal P} .
\label{eq:evolution}
\end{equation}
The terms on the left are the convective derivative, indicating that 
$\bv \cdot \bb$ is advected by the velocity field. The terms on the 
right are divergences: using that $\nabla \cdot {\bf b} = 0$, 
and when integrated over volume with the proper 
boundary conditions (e.g. periodic boundaries) they vanish. This expresses 
the simple fact that the total cross helicity is an ideal invariant in 
MHD.

However, gradients of kinetic energy and pressure gradients affect the 
{\it local} alignment 
between the two fields. The first term on the right of Eq. 
(\ref{eq:evolution}) shows that gradients in the kinetic energy 
(e.g., shear) can change the alignment between the fields when they 
are parallel to the magnetic field lines. Indeed, a magnetic field 
line (which behaves as a material line as follows from Alfv\'en's 
theorem) tends to be distorted by the shear, and aligned with the 
velocity field locally.  For a planar shear, this would be very similar 
to what is called field-line stretching. 
Pressure 
gradients aligned along magnetic field lines, from the second term on the right, also change the alignment. 
Where a pressure gradient is present, velocity goes from the region of 
higher pressure to the region of lower pressure. If the pressure 
gradient has a projection onto the magnetic field, the resulting 
velocity field will be aligned as a result with the magnetic field. 
Moreover, from dimensional analysis we can estimate the time for 
the local alignment to take place as $\sim b_l/l$, where $b_l$ is the 
amplitude of the magnetic fluctuations at scale $l$.

Note that the induction equation is formally equivalent to the vorticity 
equation in hydrodynamics. Consequently by the same reasoning, it can be shown that
the hydrodynamic 
velocity and vorticity fields tend to align locally, as found 
numerically in \cite{PelzEA85} for regions of low dissipation. 
This, replacing ${\bf b}$ by the vorticity $\mbox{\boldmath $\omega$}= \nabla \times {\bf v}$ 
in Eq. (\ref{eq:evolution}), occurs
according to alignment of $\mbox{\boldmath $\omega$}$ with gradients of the kinetic energy and the pressure.

\begin{figure}
\centerline{\includegraphics[width=6.5cm]{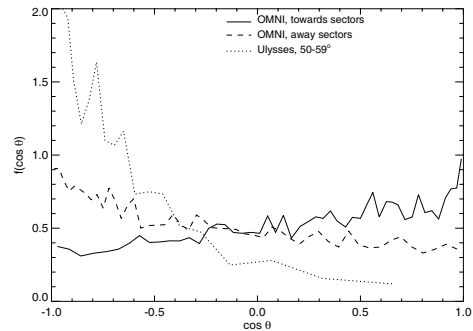}}
\caption{Pdf of $\cos(\bv,\bb)$ from 30 years of 
Omni data (ISEE, IMP and other satellite data).
Also shown is the pdf of $\cos(\bv,\bb)$ from Ulysses spacecraft data 
between 50 and 59 $^\circ$ heliospheric North latitude during a polar pass 
in solar minimum conditions. 
}
\label{fig6}
\end{figure}

{\it Solar wind observations.}
Using samples of spacecraft data we computed distributions
of the alignment angle for two interplanetary 
datasets -- the Omni dataset at 1AU near Earth orbit
in the ecliptic plane, and a sample of Ulysses data from high heliographic latitude.
Fig. (\ref{fig6}) shows the results of these analyses.
The low latitude OMNI analysis is divided into intervals in which the 
large scale interplanetary magnetic field is directed either away from or towards the sun. 
The net Alfv\'enicity is outward at the higher latitude of the Ulysses sample.
In each of these cases, the pdfs of the {\it local} alignment are 
consistent with the net cross helicity in each sample.

\begin{figure}
\centerline{\includegraphics[width=6.cm]{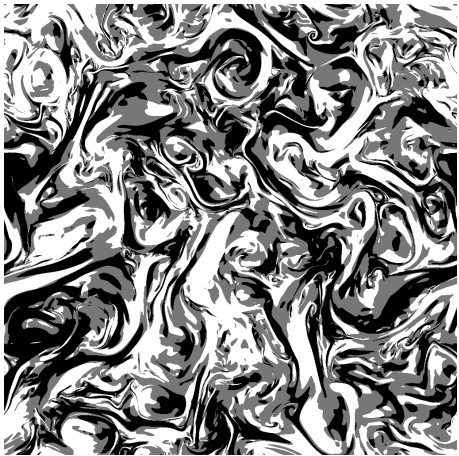}}
\caption{Cross-helicity density in a 2D incompressible MHD 
simulation, showing areas that have values of $\cos(\bv,\bb)< -0.7$ (black), 
 $|\cos(\bv,\bb)|< 0.7$ (gray) and $\cos(\bv,\bb) > 0.7$ (white).
Areas having highly aligned or anti-aligned velocity and magnetic 
field fluctuations dominate the picture. }
\label{fig8}
\end{figure}

{\it Discussion and conclusions.}
The characteristic pdfs of $\cos \theta$ 
described above cannot be explained as 
a superposition of two uncorrelated Gaussian distributions for the 
velocity and magnetic fields, although the pdfs of the velocity and 
magnetic field themselves are Gaussian (but clearly correlated).  Pdfs 
computed from random broadband uncorrelated Gaussian-component
velocity and magnetic fields have a flat $f(\cos \theta)$ distribution.
For the coherent ABC flows, $\cos \theta$ peaks at $0$ initially, 
while for the non-helical RAN1 and RAN3 flows 
the distribution is initially flat.
All cases evolve towards the characteristic shape that is high-peaked 
at $|\cos \theta| = 1$.   
In contrast, 
prior studies have shown that 
the distribution of the induced {\it emf}, ${\bf v} \times {\bf b}$,
is accurately computed from the Gaussian statistics, for 
both high and low cross helicity, in simulations and in solar wind data
 \cite{BreechEA03}. 
What apparently accounts for the difference is that the induced 
{\it emf} does not correspond to a conserved quantity, while
the alignment angle is closely associated with the ideally conserved cross helicity. 
The {\it emf} can be accounted for using Gaussian statistics, but alignment,
even of Gaussian fields, is a dynamical quantity constrained by 
the local transport and conservation, as implied by Eq. (4).

Note that Alfv\'en vortices \cite{PetviashviliPokhotelov92}, which are
coherent structures 
predicted for reduced MHD, 
have been recently observed in space plasmas \cite{SundkvistEA05};
the generalized Alfv\'en condition 
obeyed by these vortices corresponds to a local directional alignment.
Evidently this type of robust alignment 
process may be influential in a variety of space 
and astrophysical plasmas in which turbulent relaxation operates, 
as well as in the neutral fluid case.

We conclude that 
directional alignment is a rapid and robust process in turbulence.
The magnetic and velocity fields respond,
as described above, according to the local values of the 
shear and pressure gradients, essentially independently of the conditions
at remote locations, leading to local
alignment or anti-alignment; Fig. \ref{fig8} illustrates this localization or patchiness
of the directional alignment, using a 2D MHD simulation (see also \cite{sparse}).  
Since the alignment appears to be an essentially universal and rapid process, 
it would not be surprising if the coherent small scale structures in turbulence 
are associated with it. Indeed, the case shown in Fig. \ref{fig5} 
is such an example where current sheets are observed 
to have maximum alignment between the velocity and the magnetic fields \cite{1536}; similarly the local
${\bf v}$-$\mbox{\boldmath $\omega$}$ alignment may explain the slow return to full isotropy in fluid turbulence.

Research supported in part by NSF under ATM-0539995 and by NASA
under NASA NNG06GD47G. 
Computer time was provided in part by NCAR. PDM is a member of 
the Carrera del Investigador Cient\'{\i}fico of CONICET.

\newcommand{\SortNoop}[1] {}  \newcommand{\au} {{A}{U}\ }  \newcommand{\AU}
  {{A}{U}\ }  \newcommand{\MHD} {{M}{H}{D}\ }  \newcommand{\mhd} {{M}{H}{D}\ }
  \newcommand{\RMHD} {{R}{M}{H}{D}\ } \newcommand{\rmhd} {{R}{M}{H}{D}\ }
  \newcommand{\wkb} {{W}{K}{B}\ }  \newcommand{\alfven} {{A}lfv\'en\ }
  \newcommand{\Alfven} {{A}lfv\'en\ }  \newcommand{\alfvenic} {{A}lfv\'enic\ }
  \newcommand{\Alfvenic} {{A}lfv\'enic\ }

\end{document}